\documentclass[preprint,prb,aps]{revtex4}

\usepackage{graphicx}

\begin{document}

\title{Millimeter wave analysis of the dielectric properties of oil shales}
\author{John A. Scales}
\affiliation{Department of Physics, Colorado School of Mines}
\email{jscales@mines.edu}
\author{Michael Batzle}
\affiliation{Department of Geophysics, Colorado School of Mines}
\email{mbatzle@mines.edu}

\begin{abstract}
Natural sedimentation processes give rise to
fine layers in shales.  If these layers alternate between organic-rich
and organic-poor sediments, then the contrast in dielectric
properties gives rise to an effective birefringence as
the presence of hydrocarbons
suppresses the dielectric constant of the host rock.
We have measured these effects with a quasioptical millimeter
wave setup that is rapid and noncontacting.
We find that the strength of this birefringence
and the overall dielectric permittivity provide two useful 
diagnostic of the organic content of oil shales. 

\end{abstract}

\maketitle

%\section{Chemical and structural aspecs of layering induced birefringence}

A common component of sedimentary rocks, kerogens are 
macromolecular compounds that are insoluble in light organic solvents.
Kerogens are complex mixtures of organic components,
primarily the result of the decomposition of algae deposited
in fine-grained sediments under anaerobic conditions.
Estimates of the recoverable hydrocarbons from kerogen-rich
shales in Colorado alone range from 500 to 1500 billion
barrels\cite{factoid} (i.e., around $10^{14}$ liters), so the
economic importance of these oil-shales is considerable.
Kerogens also play a
role in discussions of the origin of life on Earth\cite{brocks}
as well as
being possible signatures of life beyond Earth\cite{henning}.

The presence of alternating layers of kerogen-rich and poor layers
in sedimentary rocks
has a strong influence on the degree
of electromagnetic birefringence.
In the first part this paper we describe a study in which
millimeter wave birefringence in
oil-shale samples from a single borehole is measured.
Our samples were taken from borehole \#3 in the
Roan Plateau of Western Colorado.
Each sample was previously
measured in bulk \cite{stanfield} for its oil-producing potential using
the Fischer retort method (i.e., pyrolysis) \cite{oilshale}.
We find that these two
data sets are strongly correlated.
Since the 
layering in the samples was on the order of a millimeter,
millimeter waves (MMW) are a sensitive tool for studying this effect.
Further, these measurements are rapid and completely non-contacting,
making them straightforward to implement.
To further explore the dielectric properties of shales, in the second
part of this work, 8 additional
samples had fine-scale Thermal Gravimetric Analysis (TGA) performed.
These 
samples represent a wide
variety of organic content.  The results
demonstrate that
low dielectric constant (around 4.2 or less)
is also a clear indicator of high organic content
in oil shales.

%\section{Background on MMW measurements}

At MMW frequencies, the propagation is via free space or waveguide.
Our setup, a picture of which is shown in [\onlinecite{scales_batzle_apl}],
is quasioptical.
(See the book by
Goldsmith \cite{goldsmith} for a thorough discussion of
quasioptics (QO).)
The measurements were performed with the
MMW vector network analyzer
(VNA) developed by AB Millimetre in Paris\cite{abm}.
The millimeter waves are generated by a sweepable centimeter
wave source (i.e., microwaves; in this case from 8-18 GHz).
For the lower frequency millimeter bands these
centimeter waves are harmonically multiplied by
Schottky diodes, coupled into
waveguide and eventually radiated into free space by
a scalar horn antenna.  A polyethylene lens focuses
the beam.
At higher frequencies (beyond 170 GHz) high harmonics of the
centimeter waves are phase locked to a 
separate Gunn source, which supplies the actual MMW power.
In any case, a sample
is placed
in the focal plane of the QO system.

The transmitted field is then
collected by an identical lens/horn combination, detected
by another Schottky harmonic detector and fed to a vector receiver
which mixes the centimeter waves down to more easily manageable
frequencies where the signal is digitized.
Reflected waves are also collected by the transmitting horn and
routed via a circulator and isolator to the vector receiver.
The source and receiver local oscillators are phase-locked.
The experiments described here were performed
in the W band (nominally 75-110 GHz) and the H
band (170-260 GHz); other bands
are readily accessible by changing waveguides and sources/detectors.
A more complete description of the system is given
in [\onlinecite{mmwspectroscopy}, \onlinecite{mvna:rsi}, 
\onlinecite{scales_batzle_apl}].

%\section{The measurements}

\begin{figure}
\centerline{
\includegraphics[width=16cm]{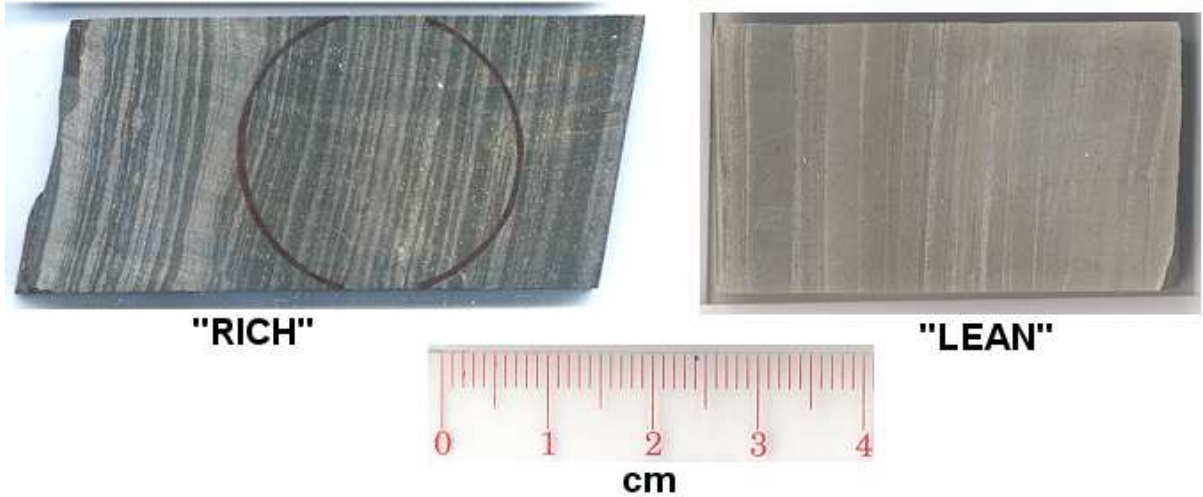}}
\caption{Examples of the  shale samples studied.  Scale is in cm.
The sample on the left has about 5 times the oil-yield potential
as the one on the right.  The geometrical layering is similar in the
two cases, but the darker color is a
clear indication of different chemical composition.  The 
two samples were about 20 m apart in situ.}
\label{608}
\end{figure}

Figure \ref{608} shows two of the five 
shale samples used in the birefringence
measurements.  On the left is shown a vertical
section through sample number 552.  The circle denotes the
size of the 
piece that was cut to fit in the optical mount. 
On the right is another sample of the same  formation
but from approximately 20 m deeper in the well.
For each of the five samples, 25 mm diameter by approximately
5 mm thick cylinders
were cut to fit into a rotational optical mount.

%\section{A uniaxial crystal model}

\begin{figure}
  \centerline{\includegraphics[width=8cm]{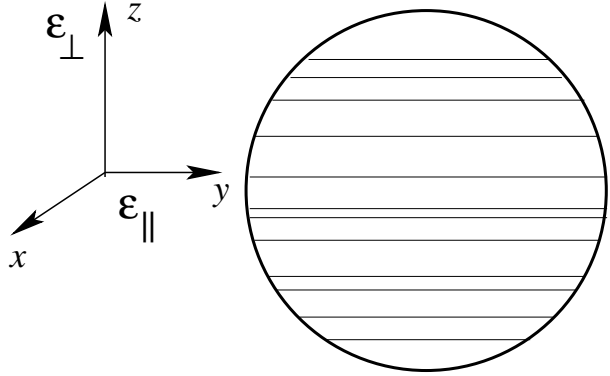}}
  \caption{We model the finely layered samples as a homogeneous, 
uniaxial
birefringent media  The optical axis is perpendicular to the layering.
In the results shown in Figure \ref{aniso}, 
$0$ degrees corresponds
to having the optical axis in the vertical ($z$) direction.}
  \label{layers}
\end{figure}

When the samples are cut in the direction of the layering,
so that the transmitted MMW see, in effect, a grating structure,
we can think of the sample as a uniaxial
crystal with optical axis perpendicular to the layering
as shown in Figure \ref{layers}.  
Uniaxial crystals 
have only two independent dielectric
tensors components\cite{ecm}.
The element along the optical ($z$) axis will be
denoted by $\epsilon _\perp$ (because the optical axis 
is perpendicular to the layering)
the elements in the $x-y$
plane are equal and will be denoted $\epsilon_\parallel$.
This is the notation used by
Goldsmith [\onlinecite{goldsmith}, page 201].  Beware that this
is the opposite of the notation used by Landau and Lifshitz in
[\onlinecite{ecm}], who regard $\epsilon_\parallel$ as being parallel
to the optical axis.

\begin{figure}
\centerline{\includegraphics[width=14cm]{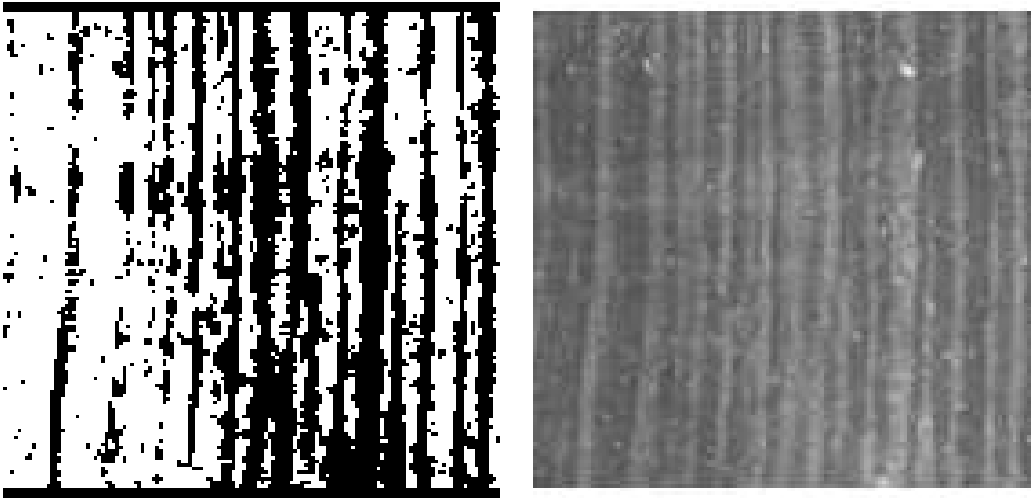}}
\caption{A 25mm by 25mm scan of the oil-rich sample 552 (right)
and a binary version (left) used for statistical analysis.
Histograms of the 
data at this threshhold level suggest that
the typical kerogen-rich layer is about 6 pixels thick 
(about 1mm) and a typical kerogen-poor layer is about 3 pixels thick
(about .5 mm).}
\label{threshhold}
\end{figure}

In our measurements 
the frequencies used (110-270 GHz) correspond to wavelengths
in the range of  2.7mm to 1.1mm in free space.  The layers are
on the order of a millimeter or less.
To get a quantitative
estimate of the spatial distribution of the layers 
we converted the optical scan in Figure \ref{608} into
zeros and ones and simply made histograms of the corresponding
distributions.  
First  
the two-dimensional (2D) scan  was converted into a binary image
(cf. Figure \ref{threshhold}).  
Then this 2D binary image was converted
to 1D by averaging along the layers.  Finally, this was 
again converted into zeros and ones
to give a binary sequence (kerogen-rich, kerogen-poor).
Then it is a simple matter to calculate the thickness of all the
layers and make histograms of these thicknesses for the two types
of material.  Given the complexity of the samples this 
procedure should be taken as a very approximate analysis.
Nevertheless, we obtain estimates of 1 mm for the average
kerogen-rich layers and .5 mm for the kerogen-poor layers.

%\section{Results}

\begin{figure}
  \centerline{\includegraphics[width=8cm,angle=-90]{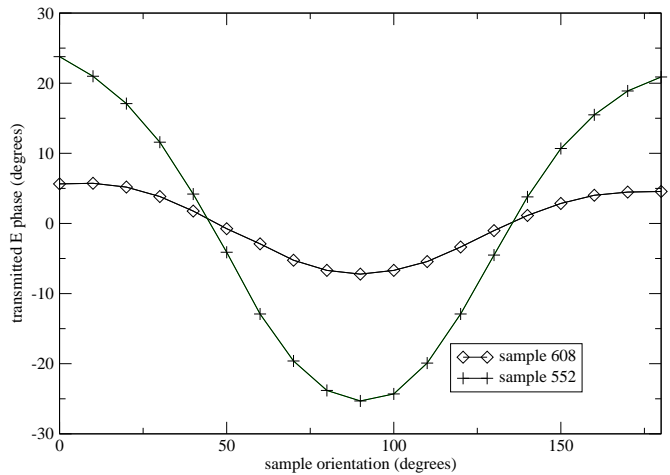}}
  \caption{Birefringence in the transmitted E field as measured
with the VNA at 110 GHz (2.7 mm wavelength). 0 degrees corresponds to
the E field parallel to the optical axis.
By previous chemical
analysis, sample 552 has 5 times the organic content of sample 608.}
  \label{aniso}
\end{figure}

Figure \ref{aniso} shows measurements of the phase of the
transmitted electric field through two of the five samples.
Our measurement uses a single-mode (TE10) Gaussian beam polarized
vertically, i.e., along the optical axis when the samples are
at 0 degrees and is normally incident on sample.
The phase of the electric field can be
read off the VNA directly.  At a given frequency there are no
apparent angle-dependent amplitude effects.

We've chosen to show the data in Figure \ref{aniso}
in terms of the phase of the E field
to emphasize the fact that this is what we measure directly.  However,
this phase also depends on the thickness of the sample, which varies 
slightly from sample to sample.  So in Figure \ref{aniso} a small
correction has been applied to remove this effect.
Henceforth, we will relate
angle-dependent phase to angle-dependent index of refraction 
(a size-independent material property) via
$\delta \phi = \frac{2 \pi d}{\lambda} \delta n$
where $\delta \phi$ is the difference in the transmitted phase 
(in radians) between the two measurements of the E field, parallel
to and perpendicular to the optical axis, $\delta n$ the corresponding
change in index, $d$ is the thickness of the sample and $\lambda$ the
wavelength.

\begin{figure}
  \centerline{\includegraphics[width=8cm]{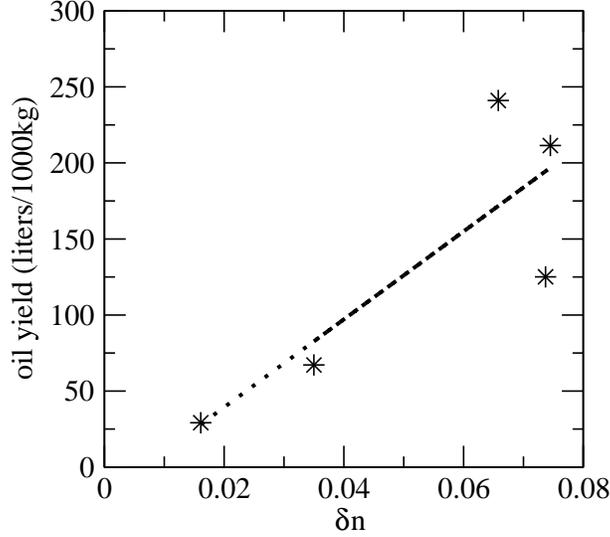}}
  \caption{Birefringence in all the oil-shale samples measured 
with the VNA at 110 GHz (2.7 mm wavelength). 
The plot shows oil-yielding potential (known from previous analysis) 
plotted against the measured birefringence in the index of refraction.
Here $\delta n$ is the difference between the index at 0 and 90 degrees
respectively.
As a point of reference, the birefringence of crystalline quartz
is about .047 at the same frequency, while for sapphire
it is .34 \cite{goldsmith}.}
  \label{kerogen}
\end{figure}

As shown in Figure \ref{aniso}, there is a strong
correlation between kerogen content and the degree of
MMW birefringence.
The possibility that we could use this birefringence to infer
organic content
led us to measure three more samples distributed though
the same borehole but with different kerogen levels.  Figure \ref{kerogen}
shows the results of MMW birefringence measurements for all five 
initial samples.
The data are plausibly consistent with
a straight line having a slope
of 28 liters/1000 kg per .01 increase in the
birefringence; the latter being
defined as the difference in the index of refraction
between 0 degrees (E polarized parallel to the optical axis) and
90 degrees (E perpendicular to the optical axis).

The kerogen itself appears to have little intrinsic birefringence.
We took the sample with relatively high kerogen
content (241 liters/1000 kg) and cut a 25mm diameter by 6.8 mm thick
cylindrical
sample so that the layers were along the axis of the cylinder
and so that the sample consisted almost entirely of the kerogen
rich shale component.
The birefringence
was .004.
This is far smaller than that 
measured for any of the samples cut with the layering perpendicular
to the optical axis.  The real part of the dielectric permittivity for this
kerogen-rich sample was 4.2 ($n=2.05$)  
The same measurement from 
one of the most kerogen-poor sample gave a permittivity of 5.4 ($n=2.3$).

\begin{figure}
  \centerline{\includegraphics[width=16cm]{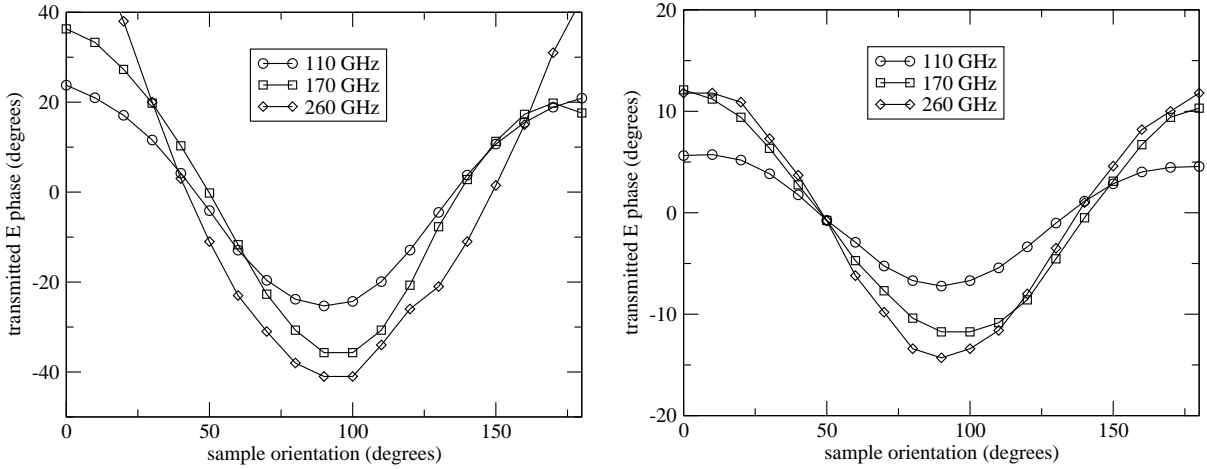}}
  \caption{Measured birefringence of the samples 552 (left) and 608 
(right) over both
W and H band.}
  \label{allfrequencies}
\end{figure}

Because the wavelengths used were comparable to the 
layer thicknesses, there is strong frequency depenence
of the birefringence.
To reach higher MMW frequencies we used a 
Gunn oscillator extension of the VNA
to span the H
band and get the birefringence in the range of 170 GHz 
to
260 GHz.  The results are shown in Figure \ref{allfrequencies}.
We can see that
the birefringence increases with frequency 
and is extremely strong at 260 GHz in the kerogen rich sample.
At 260 GHz the wavelength in free space is only 1.1 mm, while in the
shales it is about half that.  That makes the MMW 
approximately resonant with the layers.

Even in the absence of layering-induceed birefringence, the dielectric
constant itself is a good indicator of the organic content of the
shales.  Oil itself has a very low index of refraction 
(about 1.5) \cite{scales_batzle_apl}.
This is much lower than the host rock and hence kerogen rich samples will
tend have a lower permittivity than kerogen poor samples.
This can lead to counterintuitive
results for geologists, who, in the field, rely on the color of the sample
as an indicator of hydrocarbons.  Dark samples 
are assumed to have more oil.  But
our results show that this can be misleading.  For example, we looked at
samples from an outcrop that had been visually classified as having high
oil yield potential.  However, they showed no birefringence and the highest
dielectric constant of any of the samples we have measured thus far, 6.35,
which would suggest an absence of   kerogen.

\begin{figure}
  \centerline{\includegraphics[width=12cm]{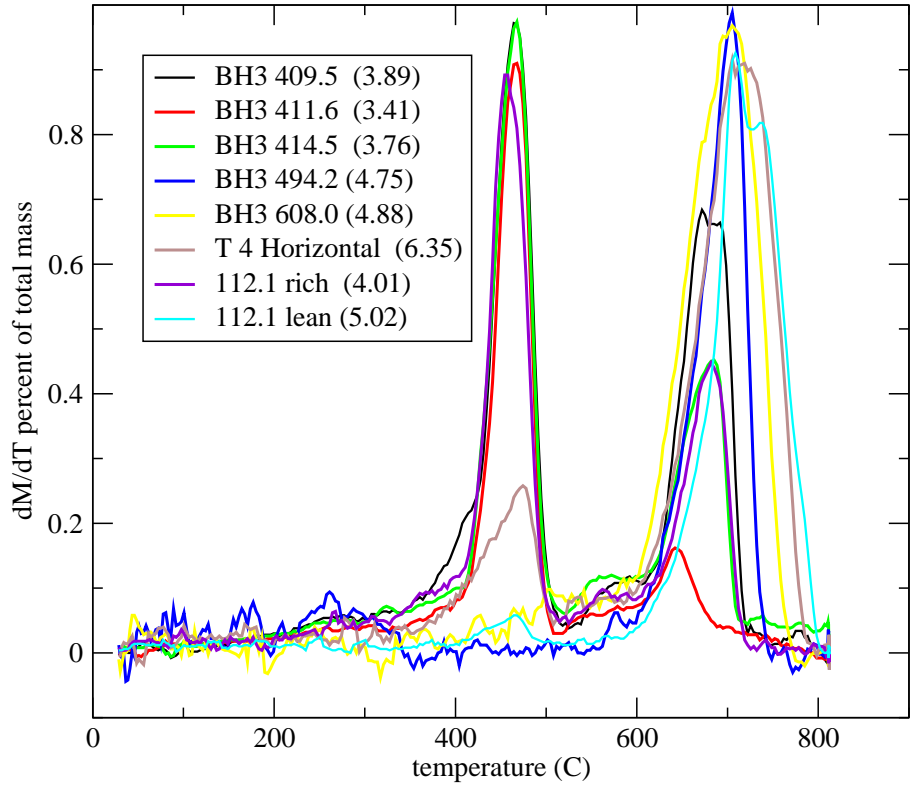}}
  \caption{Thermal gravimetric analysis (TGA) of 8 samples.  The real part
of the dielectric constant is shown in parantheses in the legends.  
For oil rich (and hence birefringent) samples, $\epsilon _ {\parallel}$ is shown.
TGA is essentially a modern retort analysis.  Small samples are heated and weighed simultaneously.  The peak at 450C is due to the breakdown of the  kerogens,
while the peak at around 700C is likely due to breakdown minerals such as 
calcite, dolomite and ankerite. \cite{shaleretort}.
 }
  \label{fig:TGA}
\end{figure}

To investigate this issue further we took
an assortment of 8
samples believed to have widely varying organic content,
measured their dielectric properties and 
performed Thermal
Gravimetric Analysis (TGA)\cite{shaleretort}.
TGA is similar to the classic retort method,
but more accurate and able to utilize much smaller samples.  
At about 450 C any hyrdocarbons will burn off giving rise to a
significant change in mass.
The results
are shown in figure
\ref{fig:TGA}.   The horizontal axis is the temperature in
Celsius and the vertical axis is the mass change with temperature ($dM/dT$)
as a percentage of the total mass of the sample.  The correlation between
low permittivity and high kerrogen yield (associated with a high peak
at   about 450C in the TGA curves) is quite clear.
In the case  of samples {\it 112.1 rich} and 
{\it 112.1 lean}, these were
thin (1-2mm) layers cut from the interior of individual layers in
a laminated sample.  The ``lean''
sample being light colored and the ``rich'' sample being dark.

In conclusion we have shown that the dielectric properties  of shales
are clear indicators of their organic content and that these properties
are easily and rapidly measured with a noncontacting millimeter wave setup.
We believe that millimeter wave dielectric spectroscopy will be a 
valuable tool in the nondestructive analysis of organic content in
natural materials.

\begin{acknowledgments}
We are grateful to Birgit Braun and
Ronny Hofman for carrying out the thermal gravimetric analysis,
and we appreicate technical discussions with
Philippe Goy of the Laboratoire Kastler Brossel,
Ecole Normale Superieure and AB Millimetre in Paris.
This work was partially supported by the National Science
Foundation (EAR-041292).
\end{acknowledgments}

\clearpage

\end{document}